\documentclass[conference, anonymous]{IEEEtran}
\IEEEoverridecommandlockouts

\usepackage{adjustbox}
\usepackage{amsmath,amssymb,amsfonts}
\usepackage{algorithmic}
\usepackage{array}
\usepackage{booktabs}
\usepackage{caption}
\usepackage{cite}
\usepackage{color}
\usepackage{enumitem}
\usepackage{graphicx}
\usepackage{listings}
\usepackage{multirow}
\usepackage{subcaption}
\usepackage{svg}
\usepackage{textcomp}
\usepackage{xcolor}

\DeclareUnicodeCharacter{0307}{\textasciitilde}

\usepackage{xspace}
\usepackage{graphicx}
\usepackage{ifthen}
\usepackage[normalem]{ulem} 
\usepackage{xcolor,colortbl}
\usepackage{hyperref}

\newboolean{showedits}
\setboolean{showedits}{true} 
\ifthenelse{\boolean{showedits}}
{
	\newcommand{\del}[1]{\textcolor{red}{\sout{#1}}} 
	\newcommand{\nbe}[3]{
		{\colorbox{#3}{\bfseries\sffamily\scriptsize\textcolor{white}{#1}}}
		{\textcolor{#3}{\sf\small$\blacktriangleright$\textit{#2}$\blacktriangleleft$}}}
}{
	\newcommand{\del}[1]{} 
	
	\newcommand{\nbe}[3]{}
}

\newboolean{showcomments}
\setboolean{showcomments}{true} 
\newcommand{\id}[1]{$-$Id: scgPaper.tex 32478 2010-04-29 09:11:32Z oscar $-$}

\ifthenelse{\boolean{showcomments}}
 {
 	\newcommand{\nbc}[3]{
 		{\colorbox{#3}{\bfseries\sffamily\scriptsize\textcolor{white}{#1}}}
		{\textcolor{#3}{\sf\small$\blacktriangleright$\textit{#2}$\blacktriangleleft$}}}
	
 }{
 	\newcommand{\nbc}[3]{}
 	
 }

\usepackage[most]{tcolorbox}
\ifthenelse{\boolean{showedits}}
{
  \newtcolorbox{inserted}{%
       title=Inserted text:,
       colframe=blue,colback=blue!5!white,
       breakable,
       leftrule=0mm, 
       bottomrule=0mm,
       rightrule=0mm,
       toprule=0mm,
       arc=0mm, outer arc=0mm,
       oversize
  }
  \newtcolorbox{deleted}{%
       title=Deleted text:,
       colframe=red,colback=red!5!white,
       breakable,
       leftrule=0mm, 
       bottomrule=0mm,
       rightrule=0mm,
       toprule=0mm,
       arc=0mm, outer arc=0mm,
       oversize
  }
  \newtcolorbox{refactored}{%
       title=Rewritten text:,
       colframe=blue,colback=red!5!white,
       breakable,
       leftrule=0mm, 
       bottomrule=0mm,
       rightrule=0mm,
       toprule=0mm,
       arc=0mm, outer arc=0mm,
       oversize
  }
}{

}
\newboolean{isblinded}
\setboolean{isblinded}{true}
\ifthenelse{\boolean{isblinded}}
{\newcommand\blind[1]{BLINDED\xspace}}
{\newcommand\blind[1]{#1\xspace}}

\newcommand{\commented}[1]{}

\newcommand{\ie}{\emph{i.e.,}\xspace}
\newcommand{\etal}{\emph{et al.}\xspace}

\definecolor{source}{gray}{0.9}

\definecolor{commentgreen}{rgb}{0.25,0.5,0.35}
\definecolor{keywordpurple}{rgb}{0.5,0,0.35}
\definecolor{backcolour}{rgb}{0.6,0.6,0.6}

\lstdefinestyle{mystyle}{
    backgroundcolor=\color{gray!5},
    commentstyle=\color{green},
    keywordstyle=\bfseries\color{purple},
    numberstyle=\tiny\color{gray},
    stringstyle=\color{black},
    basicstyle=\ttfamily\footnotesize,
    breakatwhitespace=false,    
    breaklines = true,
    breakatwhitespace=true,
    breakindent=-1ex,
    captionpos=b,                    
    keepspaces=true,                 
    numbers=none,
    numbersep=5pt,                  
    showspaces=false,                
    showstringspaces=false,
    showtabs=false,                  
    tabsize=2,
    frame=single,
    escapeinside={(*@}{@*)},  
    postbreak=\mbox{\textcolor{red}{ }\space}  
}

\begin{document}

\title{Phishing Awareness via Game-Based Learning}

\author{
\IEEEauthorblockN{Argianto Rahartomo}
\IEEEauthorblockA{
Technische Universität Clausthal\\Germany\\
\href{https://orcid.org/0000-0002-9592-0023}{ORCID: 0000-0002-592-0023}
}
\and
\IEEEauthorblockN{Ahmed Tareq Ali Ghaleb}
\IEEEauthorblockA{
Technische Universität Clausthal\\Germany\\
\href{https://orcid.org/0009-0003-9340-2980}{ORCID: 0009-0003-9340-2980}
}
\and
\IEEEauthorblockN{Mohammad Ghafari}
\IEEEauthorblockA{
Technische Universität Clausthal\\Germany\\
\href{https://orcid.org/0000-0002-1986-9668}{ORCID: 0000-0002-1986-9668}
}
}

\maketitle

\begin{abstract}
The increased use of digital devices and applications has led to a rise in phishing attacks.
We develop a serious game to raise awareness about phishing attacks and help people avoid these threats in a risk-free learning environment.
This game targets three types of phishing—clone phishing, SMS phishing, and spear phishing—and uses a Large Language Model to generate dialogues and questions dynamically.
It also incorporates state randomization and time-limited challenges to enhance the gameplay.
We evaluated two groups of participants and found that those who played the game showed, on average, a 24\% increase in awareness and a 30\% boost in confidence.
\end{abstract}


\section{Introduction}
\label{sec:introduction}

Phishing is a threat in which attackers pretend to be trusted organizations in emails or messages to steal personal information, such as usernames, passwords, and credit card details.
Typically, victims receive emails or messages that appear legitimate but point them to fake websites.
This stolen information can enable unauthorized access to financial accounts and potentially lead to identity theft.
Hence, it is important to learn how to recognize and avoid phishing.

Three common types of phishing attacks are clone phishing, SMS phishing, and spear phishing~\cite{abdillah2022}.
Clone phishing involves creating duplicates of legitimate websites to trick users into thinking they are on the original site.
Victims, often deceived by the similar appearance of the site, are tricked into entering sensitive information such as usernames, passwords, or PINs, typically through cloned login pages.
SMS phishing, also known as ``smishing'', involves attackers using SMS messages to perform phishing scams.
These messages often appear to come from reputable sources like banks or government agencies and usually urge recipients to act quickly or respond to an urgent alert.
%
%
Spear phishing targets employees within corporate or industrial entities through email campaigns.
In these attacks, deceivers use specific details to gain recipients' trust and leverage this trust to steal sensitive information or gain unauthorized system access.

Research has shown that games can significantly improve learning outcomes. 
For example,Rahartomo \etal~\cite{rahartomo2024} applied a game-based teaching approach to successfully educate dyslexic students about cryptography. 
Jayakrishnan \etal~\cite{jayakrishnan2022} developed a serious game aimed at training organizational employees to identify and avoid email phishing.
In another effort, Yasin \etal~\cite{yasin2025} created a serious game to raise awareness of clone phishing, but their approach is limited to predefined URLs, restricting its versatility.

To address the gap in tackling the human element of phishing defense~\cite{naqvi2023}, we designed a 3D game to improve phishing awareness in an engaging and risk-free environment. The game features a variety of missions in which players encounter scenarios that involve clone phishing, SMS phishing, and spear phishing, to develop their ability to recognize and respond to phishing threats.
The game leverages state randomization and time-based challenges to create a dynamic and stimulating experience. Additionally, it integrates a Large Language Model to dynamically generate dialogues and questions, providing educational insights about phishing threats and further enriching the gameplay experience.
The game is open-source,\footnote{\url{https://github.com/ahmed-ghaleb-2022/Phishing-Awareness}} and a demonstration is available online.\footnote{\url{https://youtu.be/tanRmxNA5k4}}


To assess the effectiveness of our game in improving players' awareness of phishing, we randomly assigned 28 participants to two groups: experimental and control.
Each group consisted of 14 individuals from diverse backgrounds and age ranges.
All participants completed a survey to evaluate their familiarity with phishing activities (pre-survey), revealing a general lack of awareness about phishing attempts.
Participants in the experimental group played the game and then completed a post-survey.
We then evaluated both groups using a standardized Google phishing quiz.
%

The results indicate that the experimental group achieved a 24\% higher average awareness score compared to participants who did not play the game. 
The game players also showed an increase of 30\% in confidence in detecting phishing attempts.


This paper is organized as follows.
Section~\ref{sec:game} describes the game design and its narratives.
Section~\ref{sec:experiment} presents our experiment, and Section~\ref{sec:result} reports the results.
Section~\ref{sec:threat} investigates potential threats to the validity of our study.
%
%
Finally, Section~\ref{sec:conclusion} concludes this paper.

\section{Game Design Components}
\label{sec:game}
We present our serious game components and design.
The game aims to test players' decision-making skills.
It includes the game’s dynamics, design, and narratives.
The game features basic components such as web, question, and dialogue generators, as well as scenarios that demand quick strategic thinking due to time-limited challenges.
We describe how these elements work together to train players in identifying and addressing phishing threats.

\subsection{Game Dynamics}
\begin{figure}[ht]
    \centering
    \includegraphics[width=\linewidth]{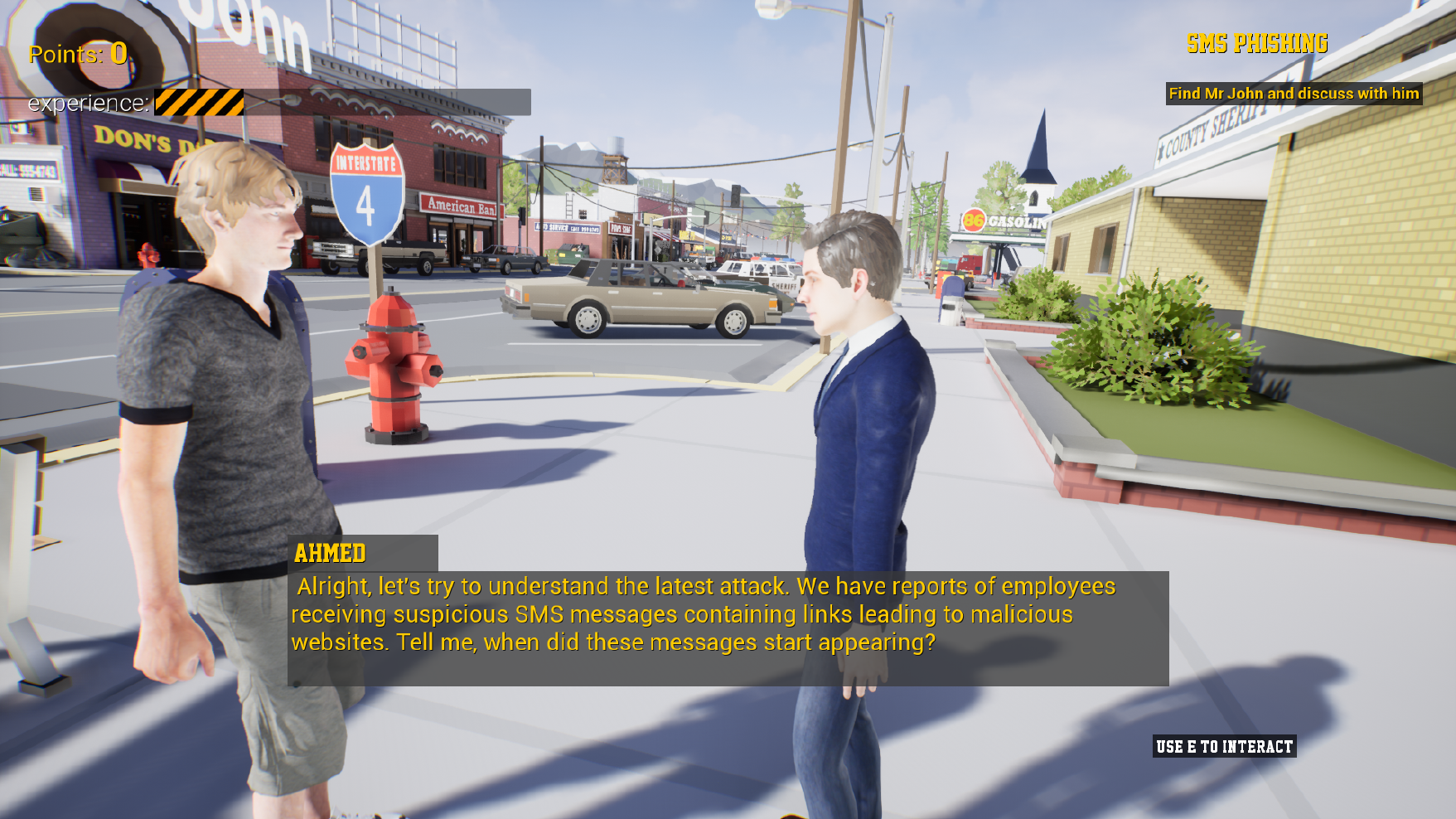}
    \caption{Player interaction with an NPC}
    \label{fig:dialogue}
 \end{figure}

\begin{figure}[ht]
    \centering
    \includegraphics[width=\linewidth]{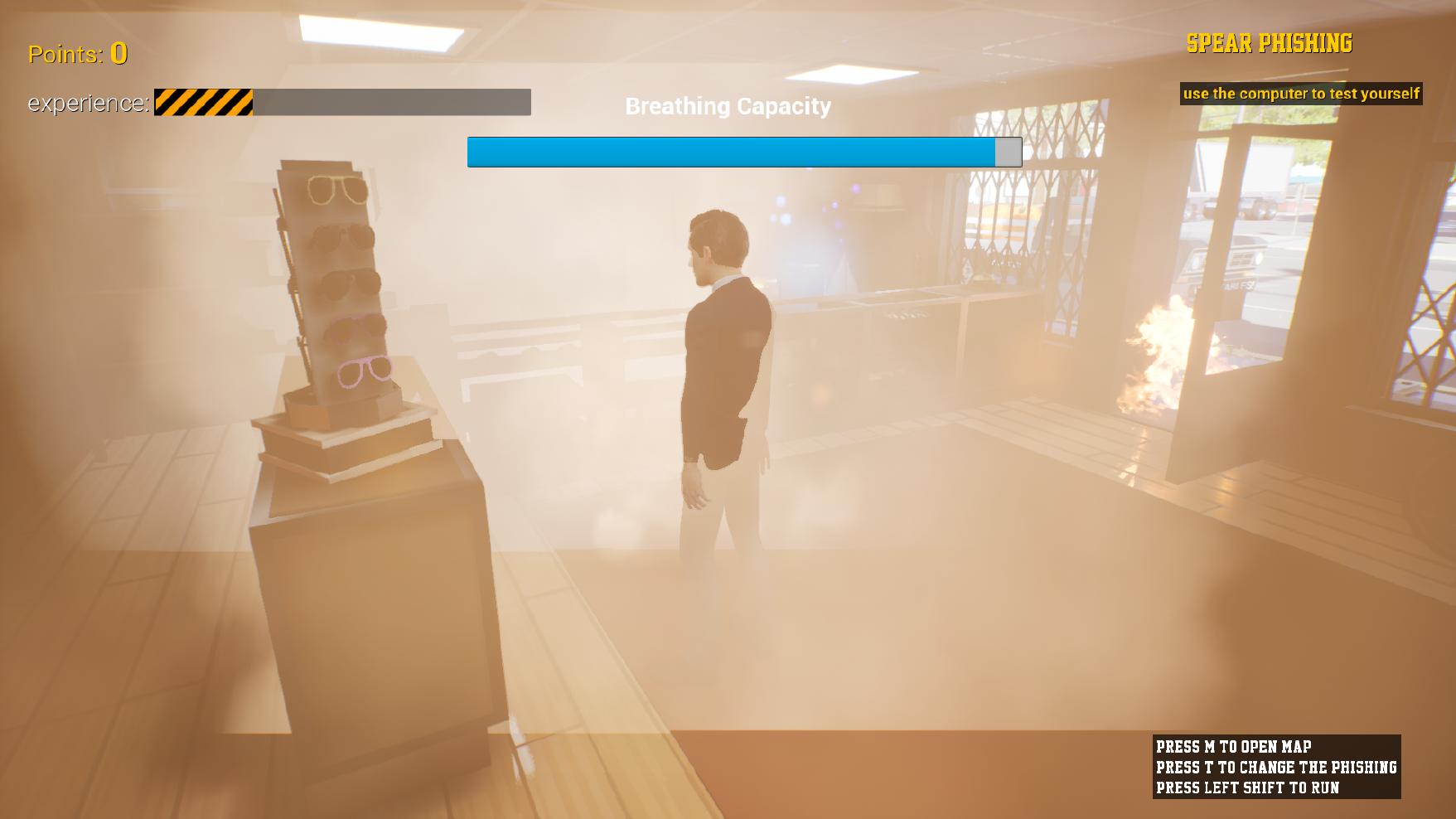}
    \caption{Gas room challenge}
    \label{fig:gasroom}
 \end{figure}

We deploy the game as an executable file produced with Unreal Engine.
It features a variety of interactive mechanics:
\begin{itemize}
    \item Multiple Choices: Players interact with non-playable characters (NPCs)—game-controlled characters—through dialogues and answer multiple-choice questions to showcase their understanding, as illustrated in Figure~\ref{fig:dialogue}.
    
    \item Decision-Making: Players check emails, SMS, and websites to determine their legitimacy.
    \item Time-Limited Challenges: Countdown timers and a gas chamber that pose lethal risks to the player’s avatar, as illustrated by the gas chamber challenge in Figure~\ref{fig:gasroom}.
\end{itemize}
%


%


\subsection{Game Design}
\begin{figure*}
    \centering
    \includegraphics[width=0.75\textwidth]{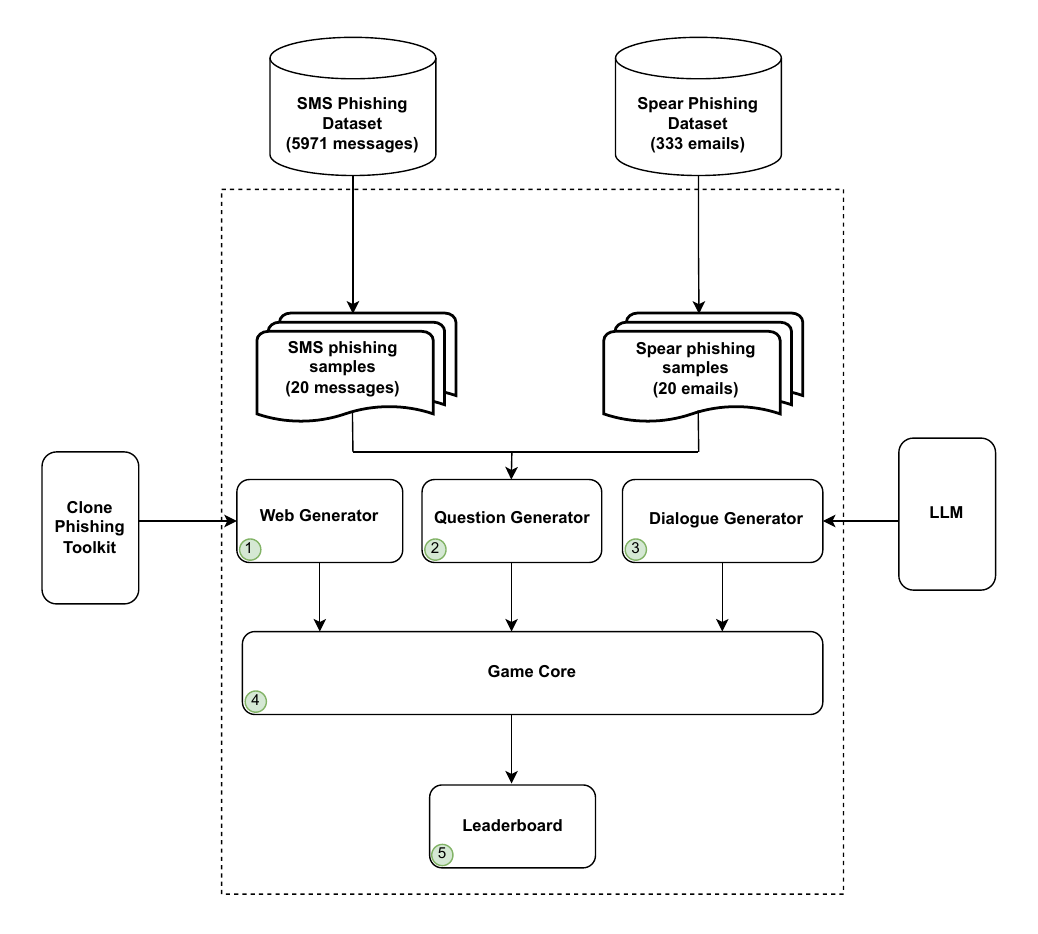}
    \caption{Phishing game building blocks}
    \label{fig:gameblock}
 \end{figure*}

Our game consists of five main components, each serving a different function.
Figure~\ref{fig:gameblock} provides a high-level overview of these components.
We build the game using a phishing toolkit, datasets, and LLM-generated dialogue.
The components include:
\begin{enumerate}
    \item \textbf{Web Generator:} This component relies on a phishing toolkit that connects through a proxy. The toolkit is based on the work of Gadient \etal~\cite{gadient2021}. This toolkit streamlines phishing website creation and facilitates proxy connections with the web generator.
    \item \textbf{Question Generator:} It uses sampled data from SMS and spear-phishing datasets to create game scenario content. We rely on a smishing dataset~\cite{mishra2023} with 5971 text messages and a spear-phishing dataset~\cite{nahmias2024} featuring 333 emails.
    The game generator randomly selects 20 samples from each dataset.
        
    \item \textbf{Dialogue Generator:} It relies on the LLM to provide dialogue at the game start. The LLM utilizes Gemma and it was hosted on GroqCloud.\footnote{https://console.groq.com/docs/models}This LLM crafts dialogues and questions for interactions between NPCs and player avatars. 
    Listing~\ref{lst:spearprompt} provides an example prompt for generating a spear phishing dialogue, while Listing~\ref{lst:spearresult} illustrates an example outcome.
    Currently, there is a need for human intervention to validate the prompt result to avoid misunderstanding and hallucination.

    \item \textbf{Game Core:} 
    We use Unreal Engine to craft our game in a 3D environment, blending educational aims with an engaging experience. It utilizes three components \ie web generator, question generator, and dialogue generator.
    \item  \textbf{Leaderboard:} It records and reports player interactions and scores.
\end{enumerate}
%
%

%

%
%
%
\begin{lstlisting}[caption={Spear phishing dialogue generator prompt}, label={lst:spearprompt}]
// Initialization
make a simple dialogue between player and NPC.

// Context setting
It is a serious game about spear phishing to increase the player skill. 

// Roles
The player is the cybersecurity expert and the NPC is the victim.

// Dialogue format
player: 
NPC: 
player: 
NPC: 
and so on
\end{lstlisting}

\begin{lstlisting}[caption={Example of a spear phishing dialogue and questions}, label={lst:spearresult}, float]
//Dialogue
player: Excuse me, but you seem to be experiencing some technical difficulties. May I offer some assistance?

NPC: Oh, thank goodness! This connection has been giving me fits. It's been happening more frequently lately, and I don't know what to (*@\textcolor{black}{do}@*).

player: I understand your frustration. Let me ask you a few questions to gather some information. Can you tell me a bit about your network infrastructure and recent security protocols?

NPC: Well, we've recently upgraded our firewall and implemented additional authentication measures, but the attacks seem to be bypassing them. It's like the attackers are constantly changing their tactics.

player: Have you noticed any specific patterns in the attacks? Time of day, sender information, or any particular themes in the phishing emails?

NPC: The emails seem very professional and convincing. They often include urgent language and threats to instill fear, making employees hesitant to question their legitimacy.

player: That suggests a skilled spearphisher. They're using social engineering techniques to manipulate your employees emotionally. We need to identify the source of the attacks and take appropriate action.

NPC: What sort of measures can we take to protect ourselves?

player: We can implement advanced email filters, train our employees to identify phishing attempts, and deploy awareness campaigns to boost our overall security posture.

(*@\textcolor{green}{//Question}@*)
Question: What is the primary purpose of the dialogue?

Option 1 (*@\textcolor{green}{(correct)}@*): To discuss the technical difficulties experienced by the NPC.

Option 2 (*@\textcolor{red}{(incorrect)}@*): To resolve the network security issues.

Option 3 (*@\textcolor{red}{(incorrect)}@*): To provide a solution for spearphishing attacks.
\end{lstlisting}

\subsection{Game Narratives}
This game is designed to improve players' abilities to recognize and avoid phishing attempts by presenting them with various scenarios that challenge their quick-thinking skills.
Players take on the role of a cybersecurity consultant who helps victims by analyzing how phishing attacks occurred and dealing with related challenges.
At the beginning of the game, players select a difficulty level that reflects the severity of real phishing incidents.
They then navigate a city using an interactive map, completing tasks to identify different types of phishing, such as clone, SMS, or spear phishing.
The map shows the real-time locations of both the player and the targets.
Gameplay includes interactive sessions with NPCs where players gather information and solve phishing-related puzzles within specific time limits.
Advancing in the game depends on completing tasks, with each correct answer earning players points and badges.

\section{Experiment Design}
\label{sec:experiment}
We outline the design of our experimental study, which aims
to evaluate the effectiveness of a serious game in improving
awareness of phishing.
%
%
We collect demographic data to ensure a diverse and representative sample.
We then proceeded with a detailed data collection phase, recording participants’ interactions with the game and gathering their feedback.
%
%
Subsequently, we assessed both groups using a standardized phishing quiz and recorded their performance scores.
The quiz comprises all three types of phishing that we cover in this game.
%

%
%
%
%
%
%

\subsection{Demography}
We conducted a survey with human participants to evaluate the game, recruiting them through postings on online forums such as Reddit and Medium, as well as through our connections.  
A total of 28 participants were recruited and evenly divided into two groups: an experimental group and a control group, each consisting of 14 participants.  
The survey included questions about participants' age, gender, and education level. Among the participants, 4 were female, and 24 were male, with ages ranging from 13 to 55 years (average age = 26.57, standard deviation = 8.53).  
The participants represented diverse fields, including IT, business administration, mechatronics, and English literature, with some still attending high school. The highest level of education achieved among them was a master’s degree.

\subsection{Data Collection}
\begin{figure}[ht]
    \centering
    \includegraphics[width=0.5\linewidth]{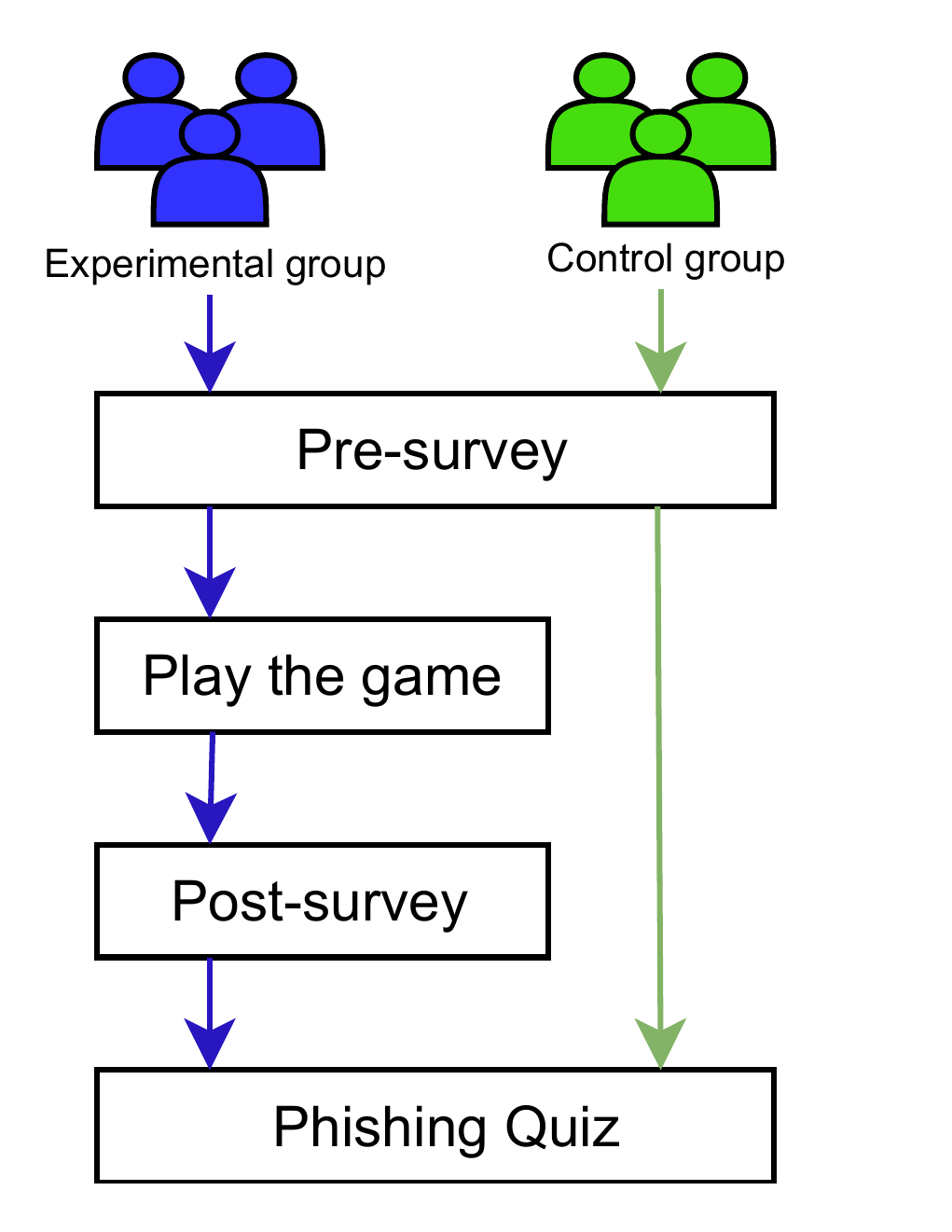}
    \caption{The experiment setup}
    \label{fig:participantflow}
\end{figure}

Figure~\ref{fig:participantflow} illustrates our experimental process.
We collect responses from participants using Google Forms through two surveys: a pre-survey before the game and a post-survey after the game, as detailed in Table~\ref{tab:pre-game} and Table~\ref{tab:post-game}.
The control group only participates in the pre-survey and not in the game or post-survey.
Their responses establish a baseline for comparing and evaluating the game's impact on the experimental group.
The experimental group first completes the pre-survey and then plays the game.
They can play the game repeatedly and without any restrictions for two weeks.
After completing the game, they fill out the post-survey to evaluate their interactions with the game and its benefits.
Finally, both groups take a standardized phishing quiz from Google, which includes ten questions.\footnote{https://phishingquiz.withgoogle.com}
We schedule the control group to take the phishing quiz one week after the experimental group.

%
%
%
%
%
%
%
%
%
%
\begin{table}[ht]
    \centering
    \caption{The pre-survey}
    \label{tab:pre-game}
    \begin{tabular}{lp{7cm}} \toprule
        \textbf{\#} & \textbf{Question} \\ \midrule
        \multirow{2}{*}{Q$_{1}$}&You are familiar with the concept of phishing.\\
        &Likert scale 1: Not at all to 5: Very familiar\\[2pt]
        \multirow{2}{*}{Q$_{2}$}&You have been a victim of a phishing attack.\\
        &Options: Yes, No, or Maybe\\[2pt]
        \multirow{2}{*}{Q$_{3}$}&You have clicked on a link or opened an attachment in a suspicious email.\\
        &Options: Yes, No, or Maybe\\[2pt]
        \multirow{2}{*}{Q$_{4}$}&You are confident in identifying phishing attempts in emails or messages.\\
        &Likert scale 1: Not at all to 5: Very confident\\
        \bottomrule 
    \end{tabular} 
\end{table}
\begin{table}[ht]
    \centering
    \caption{The post-survey}
    \label{tab:post-game}
    \begin{tabular}{lp{7cm}} \toprule
        \textbf{\#} & \textbf{Question} \\ \midrule
        \multirow{2}{*}{Q$_{1}$}&Your understanding of phishing has improved after playing the game.\\
        &Likert scale 1: Strongly disagree to 5: Strongly agree\\[2pt]
        \multirow{2}{*}{Q$_{2}$}&You are likely to recommend this game to others for learning about phishing.\\
        &Likert scale 1: Strongly disagree to 5: Strongly agree\\[2pt]
        \multirow{2}{*}{Q$_{3}$}&You are now confident in identifying phishing attempts.\\
        &Likert scale 1: Strongly disagree to 5: Strongly agree\\[2pt]
        Q$_{4}$&What did you find most helpful about the game in understanding phishing?\\[2pt]
        Q$_{5}$&Were there any parts of the game that were unclear or confusing?\\[2pt]
        Q$_{6}$&What changes, if any, would you make to your online security habits after playing the game?\\[2pt]
        Q$_{7}$&Do you have any suggestions for improving the game?
        \\\bottomrule 
    \end{tabular} 
\end{table}
%

\section{The Result}
\label{sec:result}
In this section, we discuss the findings of our study, which evaluated the impact of our serious game on participants' awareness of phishing, confidence, and ability to identify phishing attempts.
The results show differences in behavior, confidence levels, and quiz scores between the experimental and control groups.
We provide detailed analyses in three main areas: overall phishing awareness, changes in participants' confidence, and their performance on a standardized phishing quiz.
The experimental group completed a pre-survey, played the game, filled out a post-survey, and then took the phishing quiz.
On the other hand, the control group only participated in the pre-survey and the phishing quiz.
%
%
%
%

\subsection{Phishing Awareness}
The pre-survey (Table~\ref{tab:pre-game}) investigates participants' familiarity with phishing (Q$_{1}$) and their previous experiences with phishing attacks (Q$_{2}$ and Q$_{3}$).
Seventy-eight percent of participants in the experimental group indicate their willingness to change their online security behaviors.
These changes include being more cautious when verifying email addresses, avoiding unfamiliar links, and considering the implementation of two-factor authentication to secure their accounts, as shown in Table~\ref{tab:post-game} Q$_{6}$.
Some participants are unsure about the necessity of adjusting their behaviors.
Overall, this activity not only raises awareness but also encourages participants to adopt better security practices.

%
%
%
%

\subsection{Confidence Increase}
The initial survey results reveal that 53.57\% of participants lacked confidence in identifying phishing attempts (Table~\ref{tab:pre-game} Q$_{4}$).
Our game significantly improved participants' confidence levels, with the percentage of moderately confident individuals increasing to 65\% and the percentage of very confident individuals rising to 35\% after playing the game (as shown in Table~\ref{tab:post-game} Q$_{3}$).
On average, participants experience a 30\% increase in confidence.

%
%
%
%
%

\subsection{Phishing Quiz Result}
\begin{figure}[ht]
    \centering
    \includegraphics[width=0.9\linewidth]{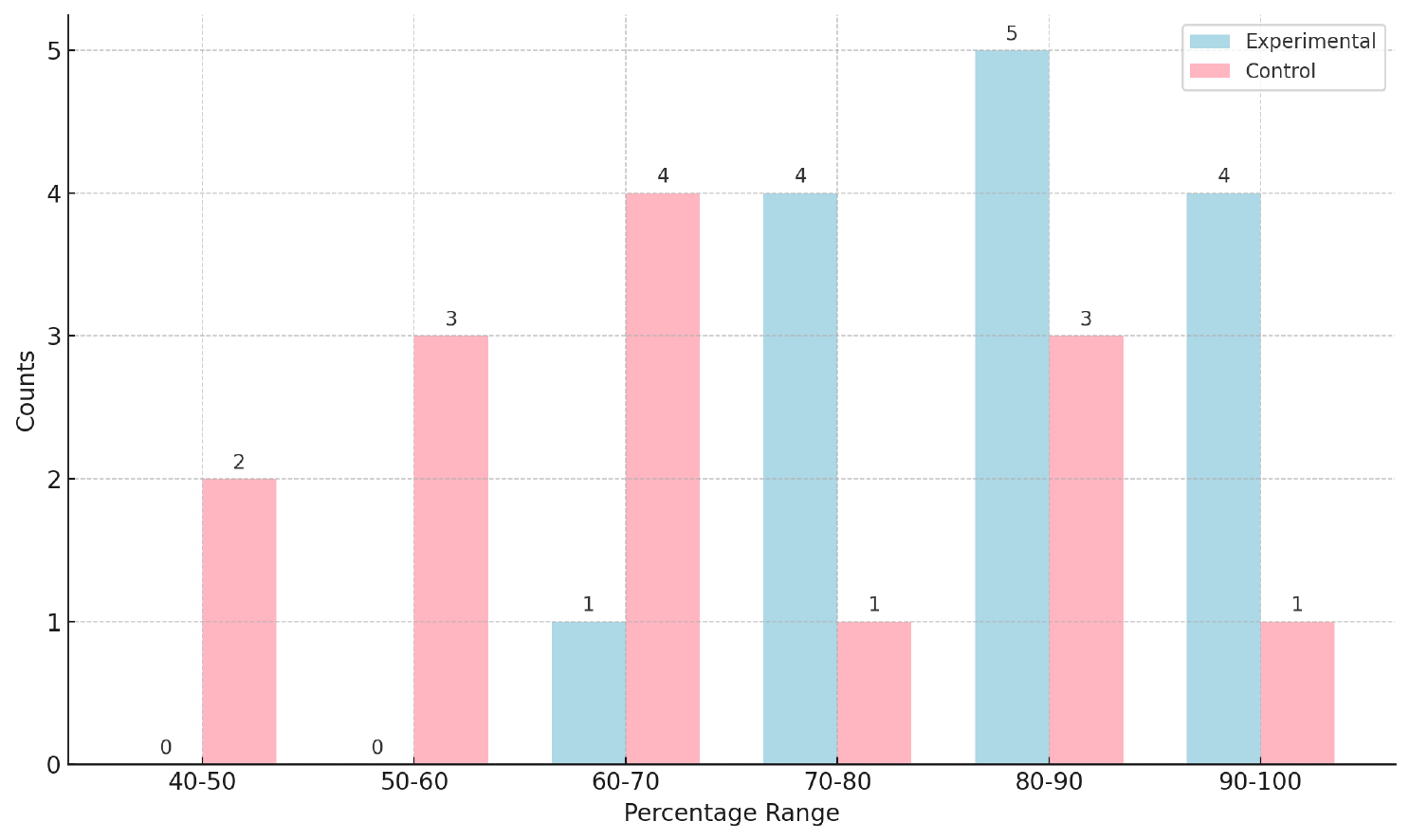}
    \caption{Score distribution among both groups}
    \label{fig:avgscoregroups}
\end{figure}

We organized participants into two groups: experimental and control.
We measured each participant's success as a percentage and sorted these percentages into 10 intervals.
We present this distribution for both groups using a bar chart, as shown in Figure~\ref{fig:avgscoregroups}.
%
%
%
The control group scored an average success rate of 62.14\% (standard deviation = 15.78\%), while the experimental group scored higher at 85.71\% (standard deviation = 11.87\%), outperforming the control group by 24\%.
The data indicate that over 90\% of the experimental group can correctly answer at least 75\% of the quiz questions, compared to only 28.57\% in the control group achieving this benchmark.

We conduct a t-test to evaluate the significance of our findings. The \emph{null hypothesis} states that ``serious games do not increase awareness'', while the alternative hypothesis suggests that they do. With a t-value of 4.46772 and a p-value of 0.00014 (which is less than 0.05), we reject the null hypothesis and conclude that our game improves phishing awareness.

%
%
%
%
%

%
%
%
%
%
%

\section{Threats to Validity}
\label{sec:threat}
The selection of participants for this study may show a bias, as volunteers often have a pre-existing interest or expertise in cybersecurity.
This could result in the findings reflecting higher-than-average knowledge or enthusiasm for phishing, which may not accurately represent the general population.
To address this, we recruit participants through various channels such as social media platforms like Reddit and Medium, universities, and community centers.

Participants may also feel social pressure to provide favorable answers, possibly overstating their vigilance or downplaying their susceptibility to phishing.
This could lead to artificially positive results.
We ensure honest responses by guaranteeing the anonymity and confidentiality of all answers.

It is necessary to ensure the reliability of the tools used to measure awareness and behavioral change.
Inconsistent quizzes or surveys might produce unreliable results, making it difficult to assess the effectiveness of the serious game in increasing phishing awareness.
Additionally, we analyze the first response from each participant to maintain data integrity.

\section{Conclusion}
\label{sec:conclusion}
We built a serious game to increase players' awareness to uncover phishing attacks in a risk-free learning environment.
We found that the game helped players to identify and mitigate malicious schemes, promoting stronger security practices.
According to the post-survey, 85\% of players gained a deeper understanding of phishing, and The game motivated 75\% of players to adopt better security practices, such as verifying email authenticity and avoiding suspicious links in their daily lives.
We also found that those who played the game showed, on average, a 24\% increase in awareness and a 30\% boost in confidence.

Future improvements could focus on refining game features, adding more scenarios, and collecting more data such as click streams.

\bibliographystyle{IEEEtran}
\bibliography{main}

\end{document}